\documentclass[12pt]{article}

\usepackage[utf8]{inputenc}
\usepackage{cancel}
\usepackage{amssymb}
\usepackage{soul}
\usepackage{tabularx}
\usepackage{xcolor}
\usepackage{booktabs}
\usepackage{subcaption} 
\usepackage{colortbl}
\usepackage{authblk}
\DeclareUnicodeCharacter{00A0}{ }

\newcolumntype{C}{>{\centering\arraybackslash}X}
\usepackage[T1]{fontenc}
\usepackage{epsfig}
\usepackage{latexsym}
\usepackage{graphicx}
\usepackage{amsmath}
\usepackage{amsfonts}   
\usepackage{amssymb}    
\usepackage{float}
\usepackage{bm}
\usepackage{url}
\usepackage{hyperref} 
\usepackage[nodisplayskipstretch]{setspace}
\def\lsim{\raise0.3ex\hbox{$\;<$\kern-0.75em\raise-1.1ex\hbox{$\sim\;$}}}
\def\gsim{\raise0.3ex\hbox{$\;>$\kern-0.75em\raise-1.1ex\hbox{$\sim\;$}}}

\def    \beq            {\begin{equation}}
\def    \eeq            {\end{equation}}
\def    \bea           {\begin{eqnarray}}
\def    \eea           {\end{eqnarray}}

\def \mn{\mu\nu{\rm SSM}}

\def\g2{{\rm GeV}^2}

\def\sw2{sin^2 \theta_w}

\def\a^tau{\alpha_{\tau}}

\def\beq{\begin{equation}}
\def\eeq{\end{equation}}
\def\beqa{\begin{eqnarray}}
\def\eeqa{\end{eqnarray}}

\newcommand{\newc}{\newcommand}
\newc\BR{BR}
\newc{\akappa}{A_{\kappa} }
\newc\deltagmtwo{\delta (g-2)_{\mu}} 
\newc\deltaamu{\Delta a_{\mu}}

\def\anti{\overline}

\def\la{\lambda}

\newc{\haa}{BR\(h_1\to a_1 a_1\)}
\newc{\abb}{BR\(a_1\to b\anti{b}\)}
\newc{\hbb}{BR\(h_1\to b\anti{b}\)}
\newc{\abund}{\Omega h^2}
\newc\bsgamma{b\rightarrow s \gamma }
\newc\bxsgamma{\overline{B}\rightarrow X_{s}\gamma}
\newc\brbsgamma{\BR(\overline{B}\rightarrow X_s\gamma)}

\allowdisplaybreaks

\title{\bf{
Electroweak superpartners scrutinized at the LHC in events with multi-leptons
}}
\author[a,b]{I\~naki Lara\thanks{inaki.lara@csic.es}}
\author[c,d]{Daniel~E.~L\'opez-Fogliani\thanks{daniel.lopez@df.uba.ar}}
\author[a,b]{Carlos~Mu\~noz\thanks{c.munoz@uam.es}} 

\affil[a]{Departamento de F\'{\i}sica Te\'{o}rica, Universidad Aut\'{o}noma de Madrid (UAM),
Campus de Cantoblanco, 28049 Madrid, Spain}
\affil[b]{Instituto de F\'{\i}sica Te\'{o}rica (IFT) UAM-CSIC, 
  Campus de Cantoblanco, 28049 Madrid, Spain}
  \affil[c]{Instituto de F\'isica de Buenos Aires UBA \& CONICET, Departamento de F\'isica,
 Facultad de Ciencia Exactas y Naturales, Universidad de Buenos Aires, 
1428 Buenos Aires, Argentina}
\affil[d]{
{Pontificia Universidad Cat\'olica Argentina, 
1107 Buenos Aires, Argentina}}

\renewcommand\Affilfont{\itshape\small}

\date{}

\begin{document}

\maketitle

\begin{abstract}
We analyze a multi-lepton signal plus missing transverse energy from neutrinos expected at the LHC for a bino-like neutralino as the lightest supersymmetric particle (LSP), when the left sneutrino is the next-to-LSP and hence a suitable source of binos. The discussion is carried out in the framework of the $\mn$, where the presence of $R$-parity violating (RPV) couplings involving right-handed neutrinos solves the $\mu$ problem and can reproduce simultaneously the neutrino data. Left sneutrinos/sleptons are pair-produced
at $pp$ collisions decaying to binos, with the latter decaying via RPV to $W\ell$ or $Z\nu$.
%
This signal can be compared with LHC searches for electroweak superpartners through chargino-neutralino production. The reduced cross section of the sneutrino/slepton production in comparison with the one of the latter process, limits the sensitivity of the searches to small sneutrino/slepton masses. 
Although the resulting compressed spectrum typically evades the aforementioned searches, we show that analyses using recursive jigsaw reconstruction are sensitive to these scenarios. 
As a by-product, we find that the region of bino masses 
$110-120$ GeV and sneutrino masses $120-140$ GeV
can give rise to a tri-lepton signal compatible with the local excess recently reported by ATLAS.
\end{abstract}


\section{Introduction}
\label{section1}

{Search for supersymmetry (SUSY) at accelerators has been focused mainly on signals with missing transverse energy (MET) inspired in $R$-parity conserving (RPC) models, such as the
minimal supersymmetric standard model (MSSM) \cite{Nilles:1983ge,Haber:1984rc,Martin:1997ns}, where significant bounds on sparticle masses have been obtained~\cite{Tanabashi:2018oca}. In particular, strongly interacting sparticles have to have masses above about 1 TeV, whereas the bound for the weakly interacting sparticles is about 100 GeV, with the exception of the bino-like neutralino which is basically not constrained due to its small pair production cross section.
Qualitatively similar results have also been obtained in the analysis of simplified 
$R$-parity violating (RPV) scenarios with trilinear lepton- or baryon-number violating 
terms~\cite{Barbier:2004ez}, assuming a single channel available for the decay of the lightest supersymmetric particle (LSP).
However, such assumption is not possible in other RPV scenarios, such as the 
the `$\mu$ from $\nu$' supersymmetric standard 
model ($\mn$)~\cite{LopezFogliani:2005yw,Escudero:2008jg},
where the several decay branching ratios (BRs) of the LSP significantly decrease the signal.
This implies that a naive extrapolation of the usual bounds on sparticle masses to the $\mn$ is not applicable. In addition, in RPV models basically all sparticles are potential candidates for LSPs, and therefore analyses of the LHC phenomenology associated to each candidate are crucial to test them.}

{
The only recent analyses~\cite{Ghosh:2017yeh,Lara:2018rwv} of signals at the LHC for LSP candidates in the
$\mn$ have been dedicated to the left sneutrino.\footnote{{The phenomenology of a neutralino LSP was analyzed in the past 
in Refs.~\cite{Ghosh:2008yh,Bartl:2009an,Ghosh:2012pq,Ghosh:2014ida}.}}
There it was shown that because the left sneutrino has several relevant decay modes, the LEP lower bound on the sneutrino mass of about 90 GeV~\cite{Abreu:1999qz,Abreu:2000pi,Achard:2001ek,Heister:2002jc,Abbiendi:2003rn,Abdallah:2003xc}
obtained under the assumption of BR one to leptons, via trilinear RPV couplings, is not applicable in the $\mn$.
The same conclusion was obtained for the left slepton, given that searches for its direct decay are not relevant because it decays through an off-shell $W$ and a sneutrino.
Thus, in Ref.~\cite{Ghosh:2017yeh} the prospects for detection of signals
with di-photon plus leptons or MET from neutrinos, and multi-leptons, 
from the pair production of left sneutrinos/sleptons and their prompt decays were analyzed.
A significant evidence is expected only in the mass range of about 100 to 300 GeV. The mass range of 45 to 100 GeV (with the lower limit imposed not to disturb the decay width of the $Z$) was covered in Ref~\cite{Lara:2018rwv}, 
studying the displaced-vertex decays of the left sneutrino LSP producing signals with di-lepton pairs.
Using the present data set of the ATLAS 8-TeV dilepton search~\cite{Aad:2015rba}, 
one is able to constrain the left sneutrino LSP only in some regions of the parameter space of the $\mn$, especially when the Yukawa couplings and mass scale of neutrinos are rather small. In order to improve the sensitivity of this search, it was proposed in~\cite{Lara:2018rwv}
an optimization of the trigger requirements exploited in ATLAS based on a high level trigger that utilizes the tracker information.
}

{
All this shows that the common lore on sparticle mass bounds must be carefully re-analyzed in the light of different theoretical models. This crucial task given the current experimental results on SUSY searches, has just started in the case of the $\mn$, and it has been concentrated for the moment on the electroweak sector as discussed above. 
In this context, where basically the masses are poorly constrained or not at all, we consider crucial to take into account
the recent searches at the LHC for events with multi-leptons plus MET~\cite{Sirunyan:2017qaj,Aaboud:2018jiw,Aaboud:2018zeb,Aaboud:2018sua}, since
they can be compared with $\mn$ signals.
This is the aim of this work.}

{In the $\mn$, when
the bino-like neutralino is the LSP with the left sneutrino the next-to-LSP (NLSP),
the RPC decays $\tilde{\nu}\to\nu\tilde{\chi}^0_1$ and\footnote{In what follows, the notation slepton (lepton) will be used for a charged slepton (charged lepton), and sneutrino (neutrino) for a neutral slepton (neutral lepton). 
The symbol $\ell$ will be used for electron, muon or tau
$\ell=e,\mu,\tau$,
and charge conjugation of fermions is to be understood where 
appropriate.} $\tilde{\ell}\to\ell\tilde{\chi}^0_1$ dominate 
over the RPV ones,
  thereby pair production of sneutrinos/sleptons at the LHC will be a source of bino pairs.
  Subsequently, binos will decay via RPV couplings to $W\ell$ or $Z\nu$, giving rise to signals with multi-leptons plus MET from neutrinos.}
  {In addition, 
we will also obtain regions of bino and sneutrino masses 
producing a tri-lepton signal compatible with the local excess reported by ATLAS~\cite{Aaboud:2018sua}. There, this signal was studied in the context of simplified RPC models, assuming wino-like chargino-neutralino production with a bino-like LSP. This scenario was further elaborated in 
Refs.~\cite{Athron:2018vxy,Carena:2018nlf}, including also dark matter constraints and the measured anomalous magnetic moment of the muon.}

The paper is organized as follows. {In Section~\ref{section0}, we will briefly review the $\mn$ and its relevant parameters for our analysis of the electroweak sector.}
In Section~\ref{section2}
we will introduce the phenomenology of the bino-like LSP with the left sneutrino as the NLSP, studying their relevant pair production at the LHC, as well as the signals.
On the way, we will analyze the decay widths, BRs and decay lengths of the bino.
In Section~\ref{section3}, we will consider 
the recent ATLAS searches for multi-leptons plus MET, and discuss
their significance on bino searches
in the $\mn$.
We will also show our
prescription for recasting the ATLAS result \cite{Aaboud:2018sua} to
the case of the sneutrino-bino scenario.
We then will show in Section~\ref{section4} the prospects for the searches at the LHC using an integrated luminosity of 100 and 300 fb$^{-1}$. Our conclusions and outlook will be presented in Section~\ref{section5}.

\section{The $\mn$}
\label{section0}

The $\mn$~\cite{LopezFogliani:2005yw,Escudero:2008jg}, is 
a natural extension of the MSSM, where the $\mu$ problem is solved and, simultaneously, the neutrino data can be 
reproduced~\cite{LopezFogliani:2005yw,Escudero:2008jg,Ghosh:2008yh,Bartl:2009an,Fidalgo:2009dm,Ghosh:2010zi}. This is obtained through the presence of trilinear 
terms in the superpotential involving right-handed neutrino superfields $\hat\nu^c_i$, which relate the origin of the $\mu$-term to the origin of neutrino masses and mixing. 
The simplest superpotential of the $\mn$~\cite{LopezFogliani:2005yw,Escudero:2008jg} is built with one 
$\hat\nu^c$:
\bea
W = &&
\epsilon_{ab} \left(
Y_{e_{ij}}
\, \hat H_d^a\, \hat L^b_i \, \hat e_j^c +
Y_{d_{ij}} 
\, 
\hat H_d^a\, \hat Q^{b}_{i} \, \hat d_{j}^{c} 
+
Y_{u_{ij}} 
\, 
\hat H_u^b\, \hat Q^{a}
\, \hat u_{j}^{c}
\right)
\nonumber\\
&+&   
\epsilon_{ab} \left(
Y_{{\nu}_{i}} 
\, \hat H_u^b\, \hat L^a_i \, \hat \nu^c 
-
\lambda \, \hat \nu^c\, \hat H_u^b \hat H_d^a
\right)
+
\frac{1}{3}
\kappa
\hat \nu^c\hat \nu^c\hat \nu^c\,,
\label{superpotential}
\eea
where the summation convention is implied on repeated indices, with  
$a,b=1,2$ $SU(2)_L$ indices
and $i,j=1,2,3$ the usual family indices of the standard model.

The simultaneous presence of the last three terms in 
Eq.~\eqref{superpotential} makes it impossible to assign $R$-parity charges consistently to the 
right-handed neutrino $\nu_R$, thus producing explicit RPV (harmless for proton decay). Note nevertheless, that in the limit
$Y_{\nu_i} \to 0$, $\hat \nu^c$ can be identified in the superpotential 
as a
pure singlet superfield without lepton number, similar to the 
next-to-MSSM (NMSSM)~\cite{Ellwanger:2009dp}, and therefore $R$ parity is restored.
Thus, the neutrino Yukawa couplings $Y_{\nu_{i}}$ are the parameters which control the amount of RPV in the $\mn$, and as a consequence
this violation is small.
 After the electroweak symmetry breaking induced by 
the soft SUSY-breaking terms of the order of TeV, and 
 with the choice of CP conservation, 
the neutral Higgses develop the vacuum expectation values (VEVs) 
$\langle H_{d,u}\rangle = \frac{v_{d,u}}{\sqrt 2}$,
the right sneutrinos 
$\langle \widetilde \nu_{R}\rangle = \frac{v_{R}}{\sqrt 2}$,
and the left sneutrinos
$\langle \widetilde \nu_{i}\rangle = \frac{v_{i}}{\sqrt 2}$,
where $v_R\sim$ TeV whereas $v_{i}\sim Y_{\nu_i} v_u\lsim 10^{-4}$ GeV because of the small contributions 
$Y_{\nu_i} \lsim 10^{-6}$
whose size is determined by the electroweak-scale 
seesaw of the $\mn$~\cite{LopezFogliani:2005yw, Escudero:2008jg}.
Note in this sense that the last term in 
Eq.~\eqref{superpotential} generates dynamically Majorana masses,
$m_{\mathcal M}={2}\kappa \frac{v_{R}}{\sqrt 2}\sim$ TeV.
On the other hand, the fifth term in the superpotential generates the $\mu$-term,
$\mu=\la \frac{v_{R}}{\sqrt 2}\sim$ TeV.

The new couplings and sneutrino VEVs in the $\mn$ induce new mixing of states.
The associated mass matrices were studied in detail in
Refs.~\cite{Escudero:2008jg,Bartl:2009an,Ghosh:2017yeh}.
Summarizing, in the case of one $\hat\nu^c$
there are 
six neutral scalars and five neutral pseudoscalars (Higgses-sneutrinos),
seven charged scalars (charged Higgses-sleptons),
five charged fermions (charged leptons-charginos), and
eight neutral fermions (neutrinos-neutralinos). 
{In our analysis of the electroweak sector below, we are mainly interested in the scalars/pseudoscalars and neutral fermions.} 

The neutral fermions have the flavor 
composition 
$(\nu_{i},\widetilde B,\widetilde W,\widetilde H_{d},\widetilde H_{u},\nu_R)$. Thus,
with the low-energy bino and wino soft masses, $M_1$ and $M_2$, of the order of TeV, and the same for $\mu$ and $m_\mathcal{M}$ as discussed above, this generalized seesaw
produces three light neutral fermions dominated by the left-handed neutrino flavor composition. {One neutrino gets its mass at tree level, whereas the other two at one loop. As discussed in Ref~\cite{Lara:2018rwv}, the tree-level mass can be approximated
as $m_{\nu} \approx {\sum_i {v_{i}^2}}/{4M}$, with
$\frac{1}{M}\equiv\frac{g'^2}{M_1} + \frac{g^2}{M_2}$.
}
The rest of neutral fermions get masses around the TeV scale.
However, 
if $M_1$ is small compared with the rest of the parameters, 
the fourth lightest eigenstate of the mass matrix, which we identify as the lightest neutralino {$\tilde{\chi}^0_1$}, is mainly bino dominated and the LSP 
with {$m_{\tilde{\chi}^0_1}\approx M_1$}, since the largest off-diagonal mass entry $m_{\tiny{\tilde B \tilde H_u}}=\frac{1}{\sqrt{2}}g'v_u$ is small.

The neutral scalars have the flavor composition 
($H_d^{\mathcal{R}}, H_u^{\mathcal{R}}, 
\widetilde\nu^{\mathcal{R}}_{R},\widetilde\nu^{\mathcal{R}}_{i}$), 
but the off-diagonal terms of the mass matrix mixing the left sneutrinos with Higgses and right sneutrinos are suppressed by $Y_{\nu}$ and $v_{iL}$, implying that the left sneutrino states will be almost pure.
The same happens for the pseudoscalar left sneutrino states $\widetilde{\nu}^{\mathcal{I}}_{i}$, which have in addition degenerate masses with the scalars 
{$m_{\widetilde{\nu}^{\mathcal{R}}_{i}}
\approx
 m_{\widetilde{\nu}^{\mathcal{I}}_{i}}
\equiv 
m_{\widetilde{\nu}_{i}}$. 
Their approximate tree-level value is
$m_{\widetilde{\nu}_{i}}^2
\approx  
\frac{Y_{{\nu}_i}v_u}{2v_{i}}v_{R}
(-\sqrt 2 A_{{\nu}_i}-\kappa v_{R}
+
\frac{\lambda v_{R}}{\tan\beta})
$,
%
%
where $A_{{\nu}_i}$ are the trilinear parameters in the soft Lagrangian,
and $\tan\beta \equiv v_u/v_d$.} As discussed in Ref.~\cite{Ghosh:2017yeh},
there is enough freedom in the $\mn$ to tune the 
soft 
parameters 
in order to get light sneutrinos.
Besides, the left sleptons will also be light, only a little heavier than the left sneutrinos
since they are in the same $SU(2)$ doublet, with the mass splitting mainly 
due to the usual small D-term contribution, $-m_W^2 \cos 2\beta$.

\section{Bino-like LSP phenomenology in the $\mn$}
\label{section2}

The pair production cross section of bino-like neutralinos at large hadron colliders is very small, since 
there is no direct coupling between the bino flavor state and the gauge bosons, and we are assuming that the rest of the spectrum remains decoupled. Binos are produced mainly through virtual $Z$ bosons in the s channel exploiting their small Higgsino flavor composition, or through the t channel interchange of virtual first generation squarks, strongly suppressed by their large masses.
Nevertheless, the bino-like LSP can be produced in the decay of other SUSY particles, which although heavier, have a higher production cross section at the LHC. That is the case when the left sneutrino is the NLSP. After production, the left sneutrinos decay to the bino LSP.


The dominant pair-production channels of sleptons at hadron colliders were studied in Refs.~\cite{Dawson:1983fw,Eichten:1984eu,delAguila:1990yw,Baer:1993ew,Baer:1997nh,Bozzi:2004qq}. The main production channels at the LHC are through a virtual $Z$ boson in the s channel for the pair production of scalar and pseudoscalar left sneutrinos 
$\tilde{\nu}\tilde{\nu}$,
a virtual $W$ boson for the production of a left slepton and a (scalar or pseudoscalar) sneutrino
$\tilde{\ell}\tilde{\nu}$,
and both virtual $Z$ and $\gamma$ for the pair production of left sleptons $\tilde{\ell}\tilde{\ell}$. Note that although the left sneutrino is lighter than its corresponding left slepton as discussed in the previous section, since the mass separation is always smaller than $m_W$, the phase space suppression makes the decay $\tilde{\ell}\to \tilde{\chi}^0_1+\ell$ dominant.

\begin{figure}[t!]
 \centering
 \includegraphics[width=218pt,keepaspectratio=true]{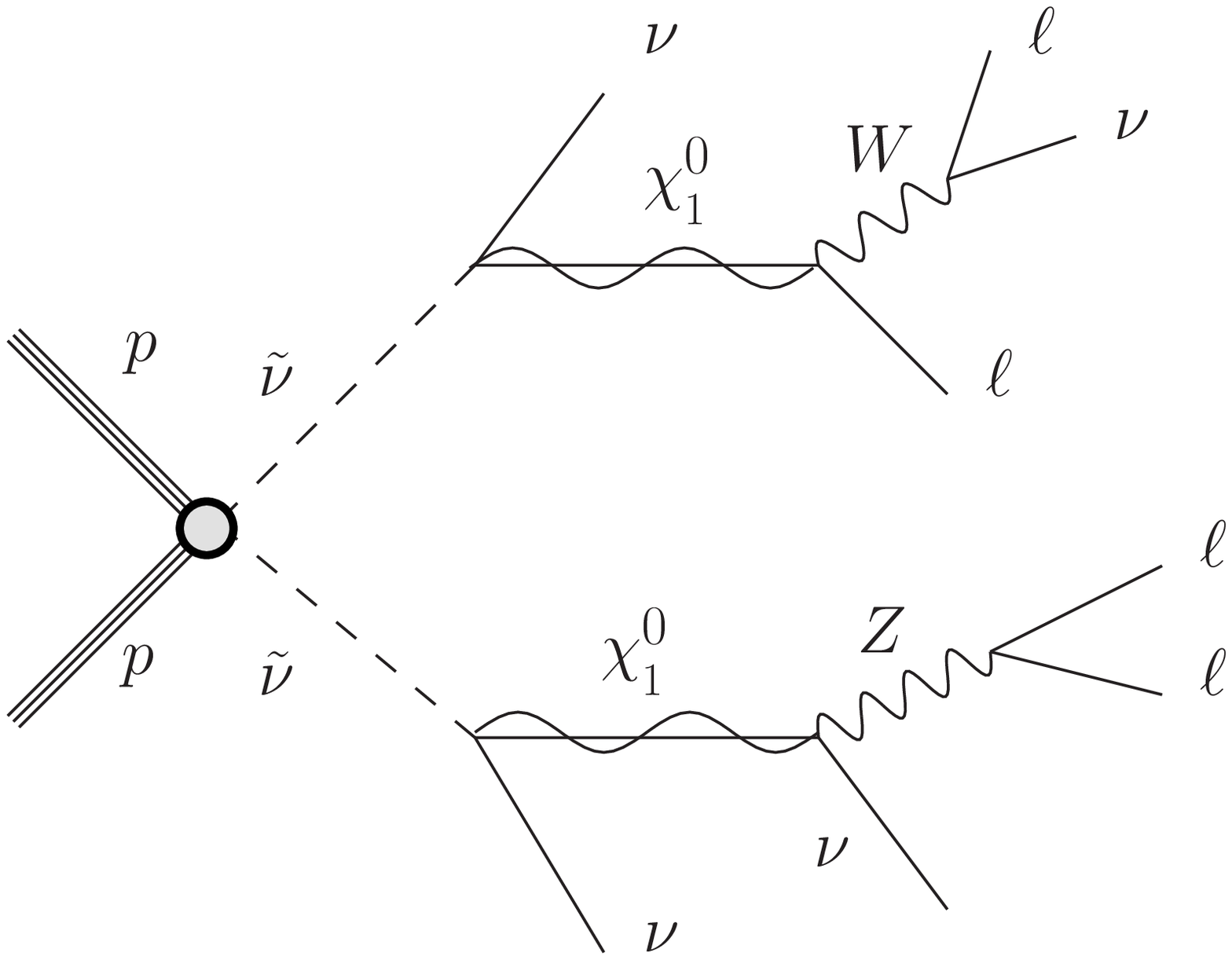}\\\vspace{30pt}
 \includegraphics[width=218pt,keepaspectratio=true]{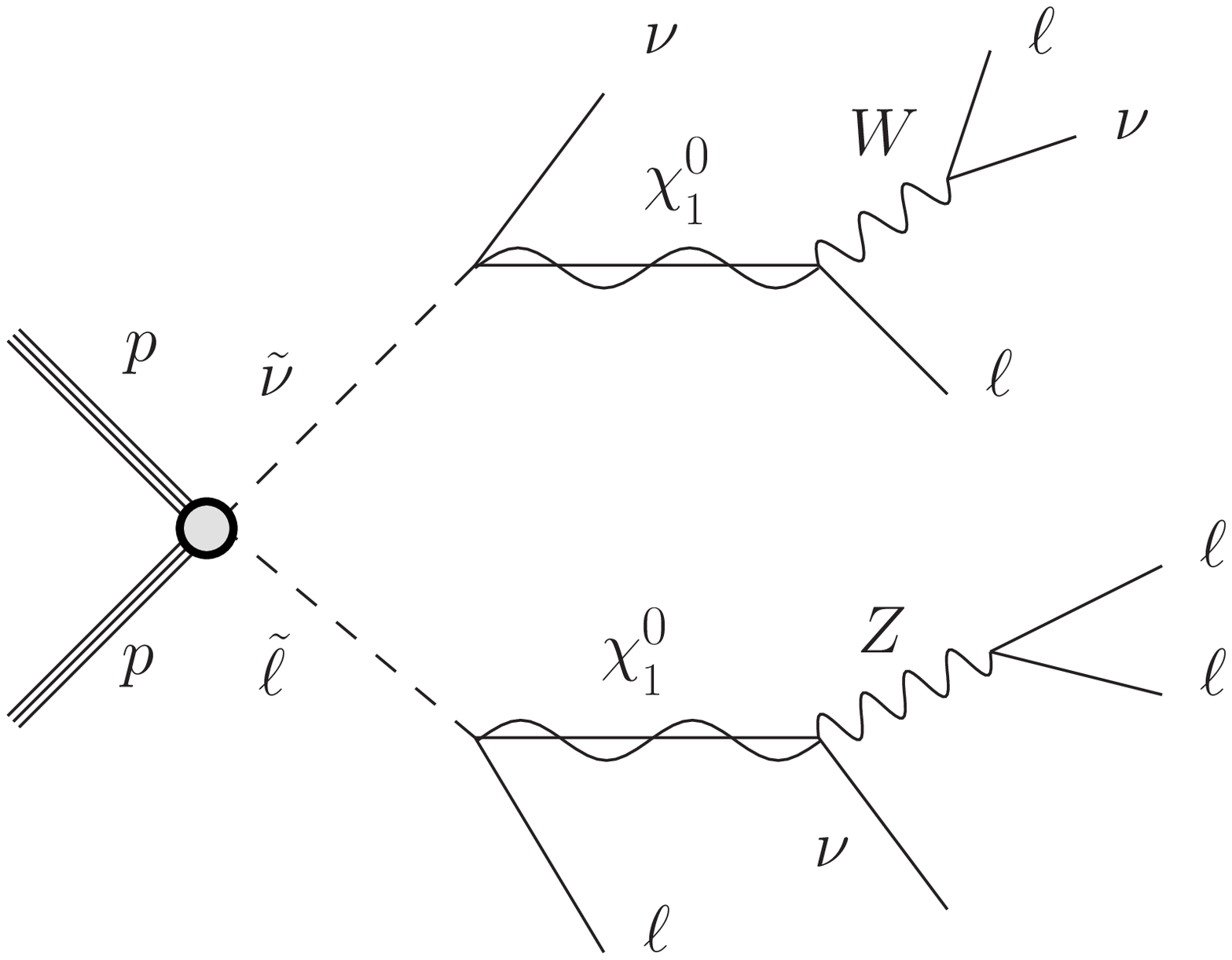}\\\vspace{30pt} 
\includegraphics
[width=218pt,keepaspectratio=true]{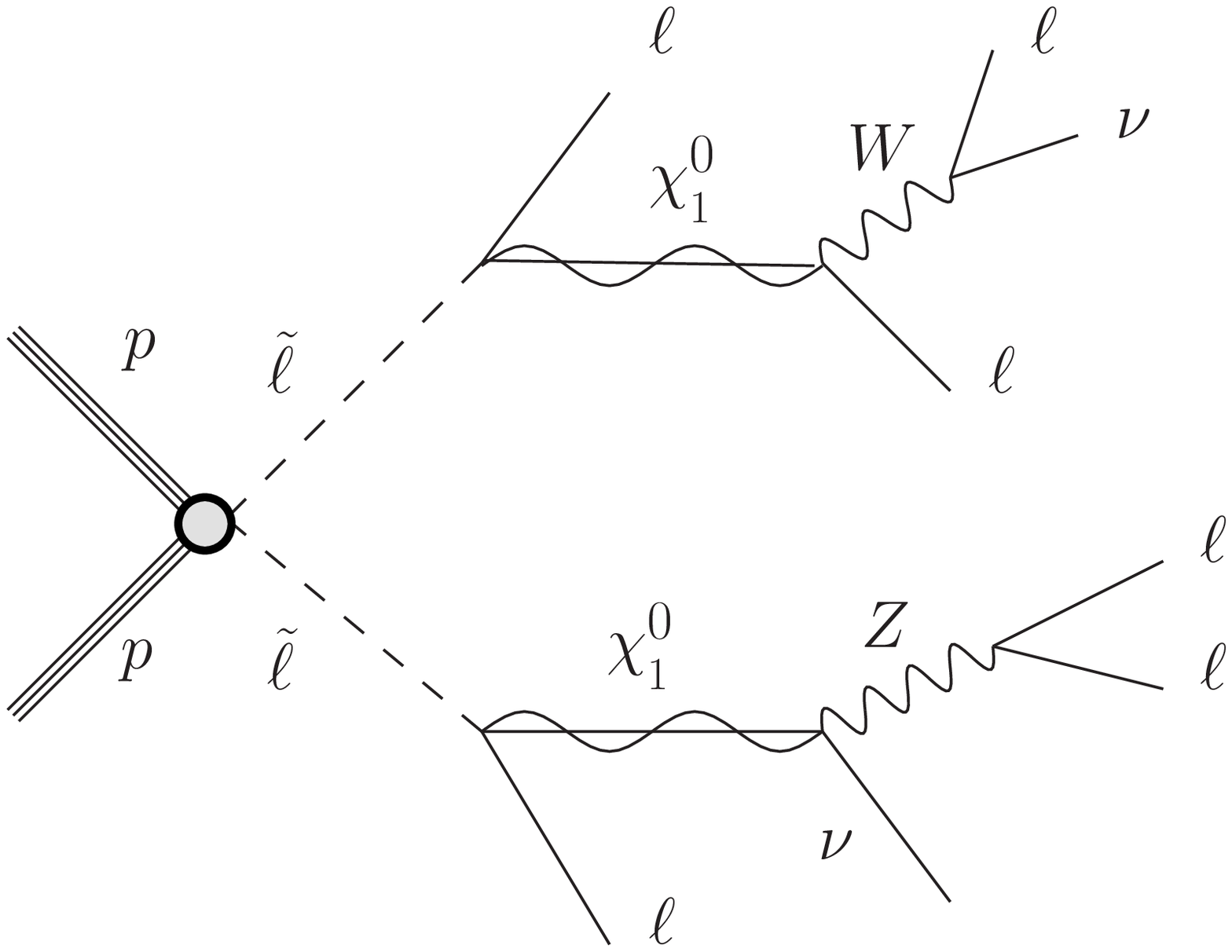}\\\vspace{20pt}
 \caption{Relevant diagrams of the benchmark $\mn$ scenario  
of RPC left sneutrino/slepton pair production, followed by the RPV decay of the 
bino-like LSP, $\tilde{\chi}^0_1$.
}
 \label{fig:production}
\end{figure}

In Fig.~\ref{fig:production}, we show the production channels as well as the RPC decays of the left sneutrino and left slepton to produce the bino LSP, {which dominate over the RPV ones since these are suppressed by the smallness of $Y_{\nu}$.}
The right sleptons can be also a source of bino LSP at the LHC. If their masses are similar to the ones of the left sleptons, an additional diagram as the third one of Fig.~\ref{fig:production} will be present. 
However, the production cross section corresponding to this extra diagram is significantly smaller than for those shown in Fig.~\ref{fig:production}. Altogether, the number of binos produced after the decay of right sleptons is around a tenth of the number produced through left sneutrinos/sleptons.

If the mass of the bino-like neutralino lies between the Higgs and $Z$ masses,
the possible two body RPV decays are to $W\ell$ and $Z\nu$, as shown also in
Fig.~\ref{fig:production}. 
There we only depicted the decay of each neutralino pair to $W$ and $Z$, and the leptonic decays of the latter. Note nevertheless that both neutralinos can also decay to $Z$ or $W$ indistinctly, and that the hadronic decay of the $W$ boson, as well as the invisible decay of one of the $Z$'s, contribute to the signal of leptonic searches, in addition to the diagrams displayed.
As we will discuss in Section~\ref{section4}, the latter processes can also contribute to the signal we are interested in analyzing, but with smaller yields.
The bino decays are mediated through the RPV mixing between the bino and neutrinos.
Although three body decays involving virtual Higgs boson or other virtual heavier scalars are possible, all of them suffer from kinematic suppression. Thus the relevant diagrams will be the two body decays, and approximate formulas for the partial decay widths are as follows:
\begin{equation}
\Gamma ({\tilde{\chi}^0\to W \ell_i})\approx\frac{g^2 m_{\tilde{\chi}^0}}{16\pi}\left(1-\frac{m^2_{W}}{m_{\tilde{\chi}^0}^2}\right)^2\left(1+\frac{m_{\tilde{\chi}^0}^2}{2m_W^2}\right)\left|U^{V}_{\tilde B\nu_i}\right|^2\,,
 \label{decay_wl}
\end{equation}
%
\begin{equation}
\sum_i \Gamma ({\tilde{\chi}^0\to Z \nu_i})\approx
\frac{g^2 m_{\tilde{\chi}^0}}{16\pi\cos^2\theta_W}\left(1-\frac{m^2_{Z}}{m_{\tilde{\chi}^0}^2}\right)^2\left(1+\frac{m_{\tilde{\chi}^0}^2}{2m_Z^2}\right)
\sum_i \left|\sum_j U^{V}_{\tilde B\nu_j}{U_{\nu_i\nu_j}^{V^*}}\right|^2\,,
 \label{decay_Znu}
\end{equation}
where $U^V$ is the matrix that diagonalizes the mass matrix for the neutral fermions~\cite{Escudero:2008jg,Ghosh:2017yeh}, and $U^{V}_{\tilde B\nu_i}$ can be approximated as 
$\frac{g' v_i}{M_1}$.

Since we are interested in the next sections in analyzing signals with electrons or muons in the final states, we focus now on the cases in our parameter space where the decay width of the neutralino to tau is smaller than to the two light families of leptons.
{This can be achieved by adjusting the sneutrino VEVs, while producing an acceptable mass scale for the neutrinos.
For example, in order to optimize the signal we can use $v_1\approx v_2>v_3$.
Then, we can compare roughly the above decay widths
summing over the two light families of leptons. 
Taking into account the kinematic factors, we obtain for the ratio an approximate
upper bound $\approx 2 \cos^2\theta_W$, implying 
that the decay to $W\ell$ will always be at least about a factor of 1.5 
larger than the decay to $Z\nu$. 
On the other hand,}
given the small value of $M_1$, from Eqs.~\eqref{decay_wl} and~\eqref{decay_Znu} we can also obtain an upper bound for the total width 
$\gtrsim 10^{-12}$ GeV, corresponding to a proper decay length $c\tau\lsim 0.2$ mm. This is short enough to expect most of the decays to happen inside the fiducial region defined by the values of the transverse impact parameter ($d^{PV}_0$) and the longitudinal impact parameter ($z^{PV}_0$) relative to the primary vertex, considered in prompt ATLAS and CMS searches. 
Subsequently, the $W$ and $Z$ bosons decay promptly producing leptons, neutrinos or jets. Thus the neutralino could be detectable in events including leptons, jets and/or MET.

Note that if the mass of the LSP drops below the mass of the $W$ boson, its decay is still possible and will proceed through three-body decays mediated by off-shell gauge bosons and scalars. The total width will be in this 
case smaller, due to the reduced phase space, and will lead to leptons and/or quarks originated at displaced vertices. This signal cannot be tested with the usual sparticles searches, but rather with dedicated analysis. The study of this possibility, although interesting, is beyond the scope of this work.

If the mass of the bino-like neutralino is larger than the one of the Higgs boson, the decay $\tilde{\chi}^0\to h \nu$ is also possible, with the dominant diagram mediated by the sneutrino-Higgs mixing. {The approximate formula for the corresponding partial decay width is given by:
 %
 \begin{equation}
{\sum_i \Gamma ({\tilde{\chi}^0\to h \nu_i})\approx \frac{g'^2 m_{\tilde{\chi}^0}}{64\pi }\sqrt{1-\left(\frac{m_{h}}{m_{\tilde{\chi}^0}}\right)^2}
\sum_i \left|\sum_j Z^H_{h \tilde{\nu}_j}{U_{\nu_i\nu_j}^{V^*}}\right|^2\,,
 \label{decay_Hnu}}
\end{equation}
where} $Z^H$ is the matrix that diagonalizes the mass matrix for the neutral scalars~\cite{Escudero:2008jg,Ghosh:2017yeh}. 
On the one hand,
the relative size of the decay of neutralino to Higgs compared to the decay to gauge bosons is suppressed by the kinematic factors.
On the other hand, the contribution of the sneutrino-Higgs mixing is not necessarily small compared to the 
bino-neutrino mixing. Moreover, 
as discussed in detail in Ref.~\cite{Ghosh:2017yeh},
the former mixing can be enhanced when the mass separation between $m_{\tilde{\nu}}$ and $m_{h^0}$ is small.
As a consequence, from the perspective of the searches 
using events with leptons, the opening of this channel softens 
the signal.

\section{Electroweak searches at the LHC}
\label{section3}

As shown in the previous section, the production and decay 
of the sneutrino (and slepton) NLSP when the neutralino is the LSP can produce signals including up to six leptons plus MET. These $\mn$ signals can be compared with searches for electroweak SUSY partners at the LHC. 

The analyses~\cite{Sirunyan:2017qaj,Aaboud:2018jiw} 
and~\cite{Aaboud:2018zeb} 
use 
the proton-proton ($pp$) collision data delivered by the LHC at a center-of-mass energy of $\sqrt s = 13$ TeV,
to search for events with two or three electrons or muons and four or more electrons, muons and taus, respectively. The former analysis 
targets direct chargino/neutralino and slepton pair production in {simplified} RPC models, whereas the latter includes the study of simplified RPV scenarios with a lepton-number violating term,
targeting direct pair production of
chargino/neutralino, left slepton/sneutrino, and gluinos.

{In the case of left sleptons/sneutrinos analyzed in Ref.~\cite{Aaboud:2018zeb} using
an integrated luminosity of 36.1 fb$^{-1}$, {the result extends the region of SUSY parameter space previously excluded by ATLAS~\cite{Aad:2014iza}} putting a lower bound of 1.06 TeV on their masses assuming a single RPV channel,
{$\lambda_{12k} \hat L_1 \hat L_2 \hat e^c_k$}, available for the decay of the neutralino LSP. {This assumption is not 
allowed} in the $\mn$ where the 
neutralino LSP has always decay channels to $W$ and $Z$ as shown in the previous section, and, in addition, the BRs of the leptonic decays of the latter are small,
of order $0.1$ and $0.03$, respectively.
Taking all this into account, we have checked that no constraint on the left sneutrino/slepton mass is obtained from this search.
}

{Concerning the searches of Refs.~\cite{Sirunyan:2017qaj} and~\cite{Aaboud:2018jiw} using 35.9 and 36.1 fb$^{-1}$, respectively, for
chargino-neutralino pair production, they assume a wino-like production mechanism 
and decays with BR one to a bino-like LSP and a leptonically decaying $W$ and $Z$ bosons into final states with three light leptons ({\it electrons or muons}) or a hadronically decaying $W$ boson and a leptonically decaying $Z$ boson into final states with two light leptons and two jets.
In this way, CMS and ATLAS obtained exclusion limits in the parameter space
$m_{\tilde{\chi}^{\pm}_1/\tilde{\chi}^0_2}-m_{\tilde{\chi}^0_1}$. For example, for a massless $\tilde{\chi}^0_1$ neutralino, 
$\tilde{\chi}^{\pm}_1/\tilde{\chi}^0_2$ masses up to approximately 600 GeV are excluded.}
This analysis uses a moderate to large amount of MET to discriminate against backgrounds, thus it is not sensitive to a compressed spectrum where this amount is not large.
Production cross sections for chargino/neutralino pairs at the LHC~\cite{Fuks:2012qx,Fuks:2013vua} are much larger than the production cross sections for slepton pairs~\cite{Fuks:2013lya}. Thus {we have obtained that} the kinematic requirement for a  mass separation between sleptons and neutralinos to have enough MET, forces the sleptons in the $\mn$ to have large masses where the expected number of pairs produced at the LCH is not enough to obtain bounds. 

A novel approach for the identification of events coming from the production of sparticles in compressed spectra, where the decay products carry low momenta, is the recursive jigsaw reconstruction (RJR)
technique~\cite{Jackson:2016mfb,Jackson:2017gcy}. This has made possible to design competitive searches for chargino-neutralino pairs even in scenarios where the mass splitting is close to the mass of the gauge bosons~\cite{Aaboud:2018sua}. As we will analyze below, the same analysis can be used to put constraints on the 
slepton/sneutrino NLSP pair production when the neutralino is the LSP in the $\mn$.

The ATLAS chargino-neutralino search using RJR in Ref.~\cite{Aaboud:2018sua} is based on the 13-TeV data with 36.1 fb$^{-1}$. All the search channels analyzed require two or 
three light leptons originated from the decay of the gauge bosons plus MET. The different signal regions are optimized to target specific mass splittings between the produced chargino-neutralino and the
neutralino LSP, for which the initial state radiation (ISR) signal regions are designed to maximize the sensitive to the case 
where $\Delta m = m_{\tilde{\chi}^{\pm}_1/\tilde{\chi}^0_2}-m_{\tilde{\chi}^0_1}$ is in the range between 100 and 160 GeV. {The results extend the region of SUSY parameter space previously excluded by ATLAS
searches in the high- and intermediate-mass regions. 
In the low-mass and ISR signal
an excess of events above the standard model prediction is observed and
the region of parameter space below $m_{\tilde{\chi}^{\pm}_1/\tilde{\chi}^0_2}=200$ GeV
cannot be excluded.
In our case,} since the production cross section of the left sneutrino/slepton is much smaller than the chargino-neutralino one, the ISR signal regions have the largest sensitivity to the mass range where the production cross section is not negligible, and $m_{\tilde{\chi}^0_1}\gtrsim m_Z$.

In the ISR signal regions, the events have to fit in the ``compressed decay tree'' described in Ref.~\cite{Aaboud:2018sua}. A signal sparticle system S decays to a set of visible momenta V and invisible momentum I recoils from a jet-radiation system ISR. The preselection criteria require exactly two or three light leptons, and between one and three non $b$-tagged jets. The transverse momentum of the leptons must fulfill $p^{\ell_{1/2}}_T>25$ and $p^{\ell_{3}}_T>20$ GeV. 
The selection criteria applied to the events after preselection are given 
in Table~\ref{Table:1}.

At first at least one same-flavor opposite sign (SFOS) pair is required, and from the formed SFOS pairs the one with invariant mass closest to $m_Z$ 
should be in the range ($75, 105$). The remaining lepton is used to construct the $W$-boson
transverse mass, $m^W_T$, as follows: $m^W_T=\sqrt{2p^\ell_TE^{miss}_T(1-\cos\Delta\phi)}$,
%
%
where $\Delta\phi$ is the azimuthal opening angle between the lepton associated with the $W$ boson and the missing transverse momentum.
{After that, the following variables are used as discriminant:}
\begin{itemize}
 \item {$p^{CM}_{T\ ISR}$: The magnitude of the vector-summed transverse momenta of the jets assigned to the ISR system.}
 \item {$p^{CM}_{T\ I}$: The magnitude of the vector-summed transverse momenta of the invisible system.}
 \item {$p^{CM}_{T}$: The magnitude of the vector-summed transverse momenta of the CM system.}
\item {$R_{ISR}\equiv \vec{p}^{CM}_I\cdot\hat{p}^{CM}_{T\ S}/p^{CM}_{T\ S}$: 
Serves as an estimate of 
$m_{\tilde{\chi}^0_1}/
m_{\tilde{\chi}^0_2/\tilde{\chi}^{\pm}_1}$.
  This corresponds to the fraction of the momentum of the system that is carried
  by its invisible system I, with momentum $\vec{p}^{CM}_I$ in the CM frame. As $p^{CM}_{T\ S}$ grows, it becomes increasingly hard for backgrounds to possess a large value in this ratio, unlike compressed signals where this feature is 
exhibited~\cite{Jackson:2016mfb}.}
 \item {$\Delta\phi^{CM}_{ISR,I}$: The azimuthal opening angle between the ISR system and the invisible system in the CM frame.}
\end{itemize}

\begin{table}[t!]
\centering
{
\footnotesize
\begin{tabular}{*8c}
\toprule
Region&$m_{\ell\ell}$ [GeV]&$m^W_T$ [GeV]&$\Delta\phi^{CM}_{ISR,I}$&$R_{ISR}$&$p^{CM}_{T\ ISR}$ [GeV]&$p^{CM}_{T\ I}$ [GeV]&$p^{CM}_{T}$ [GeV]\\
\hline
SR3$\ell\_$ISR&$\in(75,105)$&>100&>2.0&$\in(0.55,1.0)$&>100&>80&<25\\
\hline
\end{tabular} 
\begin{tabular}{*8c}
\hline
Region&$m_{Z}$ [GeV]&$m_J$ [GeV]&$\Delta\phi^{CM}_{ISR,I}$&$R_{ISR}$&$p^{CM}_{T\ ISR}$ [GeV]&$p^{CM}_{T\ I}$ [GeV]&$p^{CM}_{T}$ [GeV]\\
\hline
SR2$\ell\_$ISR&$\in(80,100)$&$\in(50,110)$&>2.8&$\in(0.4,0.75)$&>180&>100&<20\\
\bottomrule
\end{tabular} 
\caption{Selection criteria for the 3$\ell\_$ISR and 2$\ell\_$ISR signal regions. The variables are defined in Refs.~\cite{Aaboud:2018sua} and~\cite{Jackson:2016mfb}.}
\label{Table:1}
}
\end{table}

Our analysis is implemented using 
the {\tt Madanalysis v5.17}~\cite{Conte:2012fm,Conte:2014zja,Dumont:2014tja} package, and validated with simulated Monte Carlo (MC) events corresponding to the production of neutralino-chargino pairs in the context of the MSSM decaying to a neutralino LSP and leptonically decaying gauge bosons, with selected masses of $m_{\tilde{\chi}^{\pm}_1/\tilde{\chi}^0_2}=200$ and $m_{\tilde{\chi}^0_1}=100$ GeV. Ten thousand events are generated using {\tt MadGraph5\_aMC@NLO v2.6.3.2}~\cite{Alwall:2014hca} 
at leading order (LO) of perturbative QCD simulating the production of the described process with the standard model files for the MSSM. Events are then passed for showering and hadronization to {\tt PYTHIA v8.201}~\cite{Sjostrand:2006za} using the A14 tune~\cite{ATL-PHYS-PUB-2014-021}, and then to \texttt{DELPHES v3.3.3}~\cite{deFavereau:2013fsa} for detector simulation.
The results of the events selection are compared with the cutflow table provided by the ATLAS collaboration, as shown in Table~\ref{Table:2}. The first column reproduces the unweighted yields from the ATLAS analysis, the second one presents the unweighted yields from our implementation, and the last one the same yields but normalized to the number of events in {the first column}.
As can be seen from the table, the numbers agree within a $20\%$ error, thus we use this implementation to obtain the efficiency map of the ATLAS search for different masses of sneutrinos/sleptons and neutralinos.

\begin{table}[t!]
\centering
{
\footnotesize
\begin{tabular}{c|c|c|c}
\toprule
Cut applied&ATLAS yield& Implemented yield& Normalized yield\\\hline
Trigger matching $\&$ Preselection&1829&1398&1829\\
$m_{\ell\ell}\in(75,105)$ GeV $\&\ m^W_{T}>100$ GeV &533&406&531\\
$\Delta\phi^{CM}_{ISR,I}>2.0$&408&308&403\\
$R_{ISR}\in(0.55,1.0)$&157&179&234\\
$p^{CM}_{T\ ISR}>100$ GeV&115&132&173\\
$p^{CM}_{T\ I}>80$ GeV&114&115&150\\
$p^{CM}_{T}<25$ GeV&73&68&89\\
\bottomrule
\end{tabular} 
\caption{Comparison between the ATLAS cutflow shown in the auxiliary figures of
Ref.~\cite{Aaboud:2018sua} and our implementation. }
\label{Table:2}
}
\end{table}

The sneutrino/slepton pair production is simulated in a similar way, but with model files generated using a suitable modified version of {\tt SARAH} code \cite{Staub:2008uz,Staub:2011dp,Staub:2013tta}, and the spectrum is generated using {\tt SPheno} {{v}}3.3.6 code \cite{Porod:2003um,Porod:2011nf}. Cross sections are calculated at NLO+NLL using {\tt Resummino v2.01}~\cite{Buckley:2014ana,Bozzi:2006fw,Bozzi:2007qr,Bozzi:2007tea,Fuks:2013vua}.
For each selected point, ten thousand MC events are generated as explained and passed through the described selection criteria. The results are then compared with the expected 
($S^{95}_{exp}$) 
and observed 
($S^{95}_{obs}$) 
upper limits obtained in the ATLAS search. 

\section{Results}
\label{section4}

{We have already obtained that results~\cite{Sirunyan:2017qaj,Aaboud:2018jiw} 
and~\cite{Aaboud:2018zeb} do not produce bounds on the bino-like neutralino and left sneutrino/slepton masses in the $\mn$.}
By using now the method described in the previous section, we calculate the current and potential limits on the
two-dimensional parameter space 
$m_{\tilde\nu}-m_{\tilde{\chi}^0_1}$ from the 
36.1 fb$^{-1}$ 
ATLAS result~\cite{Aaboud:2018sua}, and discuss the prospects for the 100 and 300 fb$^{-1}$ searches.

The points analyzed in the $\mn$ parameter space show all a worse efficiency passing the selection requirements of SR2$\ell\_$ISR in comparison with SR3$\ell\_$ISR
in Table~\ref{Table:1}. Thus the results discussed here are derived from the limits corresponding to the SR3$\ell\_$ISR signal region.
Note first that the preselection requirement of exactly three light leptons, implies that 
the separation of masses between slepton/sneutrino and neutralino in 
the processes shown in Fig.~\ref{fig:production} must be small enough to avoid the 
leptons produced in the decay of sleptons to contribute to the signal. Otherwise, these processes are discarded from the ATLAS searches.
The processes in Fig.~\ref{fig:production}
show the highest yield of the possible combinations of neutralino decays.
Other possibilities, like $W$ decaying hadronically or both neutralinos decaying to $Z$ bosons, with only one of them decaying leptonically, contribute also to the signal, but with smaller yields. The possibility of both neutralinos decaying to $Z$ bosons, with the latter
decaying to leptons, produce a negligible contribution caused both by the small corresponding BR and the excess of predicted signal leptons.

\begin{table}[t!]
\centering
{
\footnotesize
\begin{tabular}{*6c}
\toprule
$m_{\tilde{\chi}^0}$&120 GeV&$m_{\tilde{\nu}}$&125 GeV&$m_{\tilde{\ell}}$&
145 GeV\\
\hline
BR$(\tilde{\ell}_{i}\to {\ell}_i \tilde{\chi}^0)$&1&BR$(\tilde{\nu_i}\to \nu \tilde{\chi}^0)$&1&BR$(\tilde{\chi}^0\to W e/\mu)$&$3.5\times10^{-1}$\\
BR$(\tilde{\chi}^0\to W \tau)$&$2.9\times10^{-2}$&BR$(\tilde{\chi}^0\to 
Z \nu)$&$2.4\times10^{-1}$&$\Gamma_{\tilde{\chi}^0}$&$1.28\times10^{-12}$ GeV\\
\hline
$\epsilon_{W_{\ell}Z_{\ell}}$&0.0092&$\epsilon_{W_{h}Z_{\ell}}$&0.0021&$\epsilon_{Z_{\ell}Z_\nu}$&0.0215\\
\hline
$\sigma(pp\to\tilde{\nu}\tilde{\nu})$&143.75 fb&$\sigma(pp\to\tilde{\nu}\tilde{\ell})$&276.32 fb&$\sigma(pp\to\tilde{\ell}\tilde{\ell})$&80.94 fb\\
\hline
\multicolumn{6}{c}{
Events above background in S$\ell3\_$ISR: 5.1 }\\
\bottomrule
\end{tabular} 
\caption{Benchmark point of the $\mn$ in the S$\ell3\_$ISR signal region.
{In the third row the efficiency $\epsilon$ passing the selection requirements is shown.}}
\label{Table:3}
}
\end{table}

Our result in the mass regions considered is that no points 
of the $\mn$ can be excluded from current data. 
{Concerning the excess of events mentioned in the previous section, let us point out} that
the observed limit in the 3$\ell\_$ISR signal region {($S^{95}_{obs}{=15.3}$)} is significantly larger than the expected limit {($S^{95}_{exp}{=6.9}^{+3.1}_{-2.2}$)}, due to a 3.02 sigma excess~\cite{Aaboud:2018sua}. {We have checked that} points of our parameter space in the region $m_{\tilde{\chi}^0_1}\in(110,120)$ and $m_{\tilde{\nu}}\in(120,140)$ can 
predict a number of events similar to the observed excess. As an example, Table~\ref{Table:3} shows a benchmark point (BP) in S$\ell3\_$ISR {with $M_1=127$ GeV, and
$\frac{v_{1,2}}{\sqrt 2}= 5.2\times 10^{-4}$ GeV, $\frac{v_{3}}{\sqrt 2}= 1.07\times 10^{-4}$ GeV.
These VEVs have typical values fulfilling the minimizations conditions, and 
producing the
heavier neutrino mass in the experimentally constrained range
$m_\nu \sim$ (0.05, 0.23) eV~\cite{An:2015rpe,Ade:2015xua},
given the neutrino mass formula discussed 
in Section~\ref{section0}. In particular, this BP predicts 5.1 events above background.
Other parameters, whose effect on the bino-like neutralino decay properties is less
significant, as can be understood from formulas in Section~\ref{section2}, are set to be $\lambda=0.2$, $\kappa=0.3$, $\tan\beta=10$, $\frac{v_R}{\sqrt 2}= 1350$ GeV, and
$M_2=1$ TeV,
throughout this computation. 
In order to optimize the signal} 
we have assumed that the masses of the three families of left sneutrinos/sleptons 
are degenerated, and therefore all of them contribute to the signal.
{Given the 
sneutrino mass formula in Section~\ref{section0},  
there is enough freedom to tune the masses in that way allowing for non-universality of the
parameters $Y_{\nu_i}$ or $A_{{\nu}_i}$. 
For this BP we set the typical values $Y_{\nu_{1,2}}=5\times 10^{-7}$,
$Y_{\nu_{3}}=1\times 10^{-7}$ and universal $A_{{\nu}}=-396$ GeV.
}

\begin{figure}[t!]
 \centering
 \includegraphics[scale=0.6,keepaspectratio=true]{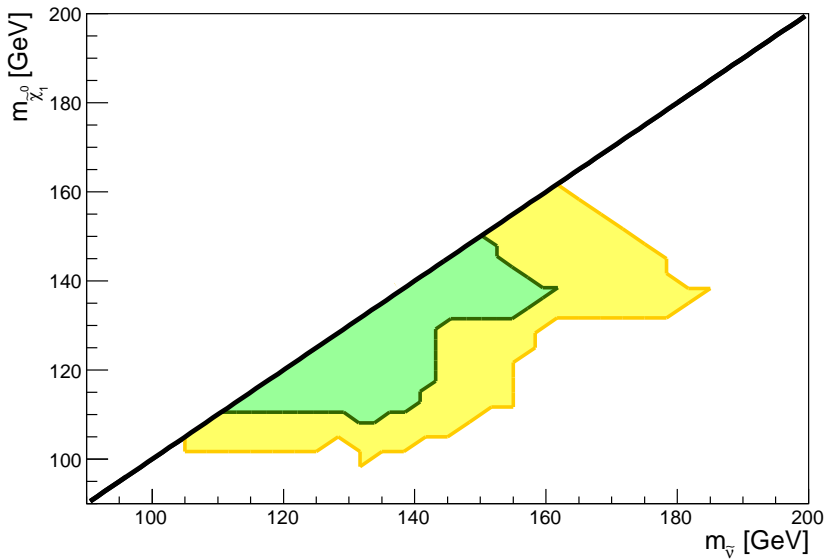}
\caption{Regions of the $\mn$ that will be probed by the signal with
three light leptons plus MET discussed in the text in the parameter space 
$m_{\tilde\nu}-m_{\tilde{\chi}^0_1}$ with universal sneutrino masses, for the 13-TeV search with an integrated luminosity
of 100 fb$^{-1}$ (green) and 300 fb$^{-1}$ (yellow).
}
 \label{fig:exclusions}
\end{figure}

If the observed local excess were due to a statistical fluctuation, and the observed upper limit converged to the expected limit, we can easily infer the potential bounds on the parameter space of the sneutrino-neutralino mass in the $\mn$. For 100 and 300 fb$^{-1}$ to be reached at the end of Run 2, we just
have to rescale the limits by 
$\sqrt{100/36.1}$ and $\sqrt{300/36.1}$, respectively. 
The result is shown in Fig.~\ref{fig:exclusions}.
{In this plot the parameters for the BP above are used, but varying $M_1$ between 95 and 215 GeV, and $A_{{\nu}}$ between $-470$ and $-376$ GeV in order to obtain the
range of neutralino and sneutrino masses shown.}
There, the solid black line shows the points where the sneutrino and neutralino are degenerated in mass. The green area encloses the excluded region for 100 fb$^{-1}$ if no excess is observed, and the yellow area shows the same for 300 fb$^{-1}$.
As we can see, the prospects show a potential exclusion of sneutrino masses up to 160 and 185 GeV for 100 and 300 fb$^{-1}$, respectively.
This region extends up to the solid line, reflecting the fact that the search will be fully sensitive to the degenerated scenario. 

In Fig.~\ref{fig:exclusions},
for a given sneutrino mass, smaller value of the neutralino mass does not give rise to a worse sensitivity until the kinematic functions in Eq.~\eqref{decay_Znu} make the product $BR(\tilde{\chi}^0\to Z\nu) BR(\tilde{\chi}^0\to W\ell)$ too small and the search becomes ineffective. 
On the other hand, for a given neutralino mass the larger the value of the sneutrino mass the smaller the production cross section becomes, limiting the exclusion scope.
{Notice also that in the lower-right corner of the figure, the limits are weaker due to the 
larger mass separation between sneutrino and neutralino. As discussed above, the increased energy of the leptons from the decay of sleptons make them to contribute to the signal giving rise to a signature with more than three light leptons.}

\begin{figure}[t!]
 \centering
 \includegraphics[scale=0.6,keepaspectratio=true]{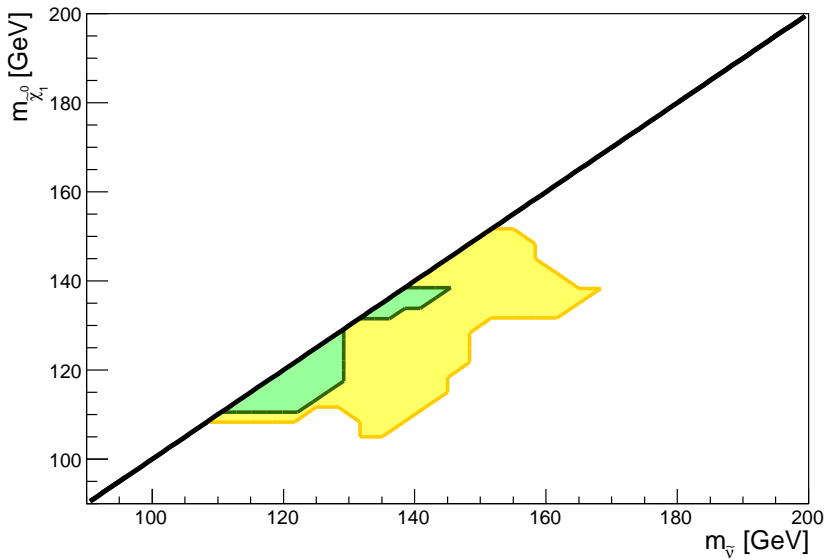}
\caption{{The same as in Fig.~\ref{fig:exclusions}, but for non-universal sneutrino masses as discussed in the text.}
}
 \label{fig:exclusions2}
\end{figure}

{To discuss how general is this result in the parameter space of the $\mn$, we can consider a (non-optimized) signal with non-degenerate sneutrino/slepton masses.
For that we can use universal $Y_{\nu_{1,2,3}}=5\times 10^{-7}$, producing 
$m_{\tilde\nu_{3}}=2.24\ m_{\tilde\nu_{1,2}}$.
Then, only two families of sneutrinos/sleptons contribute to the signal, and the result is shown in Fig.~\ref{fig:exclusions2} with $m_{\nu}=m_{\tilde\nu_{1,2}}$. Obviously, the limits are less stringent, but still some regions of the parameter space can be bounded. The prospects show a potential exclusion of sneutrino masses up to 145 and 170 GeV for 100 and 300 fb$^{-1}$, respectively.}
{A similar result would be obtained assuming universality of VEVs and masses,
since on the one hand the latter assumption increases the signal but on the other hand the former decreases it.}

\section{Conclusions and outlook}
\label{section5}

{
There is a lack of experimental bounds on the masses of the electroweak superpartners in the $\mn$ from accelerator searches, apart from very specific regions of a sneutrino LSP with a mass between 45 and 100 GeV~\cite{Lara:2018rwv}.
In order to fill this gap in SUSY searches, it is crucial to analyze all recent results that can potentially lead to (even small) limits on sparticle masses in this model.
Following this line of thought, we have discussed in this work how stringent are for the parameter space of the $\mn$ the
recent LHC searches for electroweak superpartners at the LCH in events with multi-leptons plus MET~\cite{Sirunyan:2017qaj,Aaboud:2018jiw,Aaboud:2018zeb,Aaboud:2018sua}.}

{We have obtained first that the results of Refs.~\cite{Sirunyan:2017qaj,Aaboud:2018jiw} 
and~\cite{Aaboud:2018zeb} do not produce bounds on the bino-like neutralino and left sneutrino/slepton masses in the $\mn$.
Concerning~\cite{Aaboud:2018sua}, this search uses the RJR technique and is therefore sensitive to three leptons of the two light families produced by a compressed spectrum of electroweak superpartners.
This type of signal is produced in the $\mn$ with a bino-like neutralino LSP when the left sneutrino is the NLSP. However,
we have obtained that no points of the parameter space of the $\mn$ can be excluded yet.
In the region of bino (sneutrino) masses $110-120$ ($120-140$) GeV we have found a tri-lepton signal compatible with the local excess reported by ATLAS~\cite{Aaboud:2018sua}.
If this excess were due to a statistical fluctuation, the 
prospects for the bounds on the parameter space of the sneutrino-bino mass in the $\mn$ 
are shown in 
Figs.~\ref{fig:exclusions} and~\ref{fig:exclusions2}. 
}

These limits can be complemented in the future by searches for displaced decays of the bino-like neutralino when its mass is below the threshold of the $W$ mass. 
In this case, three-body decays mediated by off-shell gauge bosons and scalars will produced a small total width due to the reduced phase space, leading
to signals with lepton and/or quarks originated at displaced vertices. 

In this work we have focused on the bino-like LSP, but another interesting possibility would be to study the case of a wino-like LSP. The different couplings involved as well as new RPV decays could modify the sensitivity of the searches to the new compressed spectrum. We plan to carry out this analysis in a forthcoming publication~\cite{prepar}.


Let us finally remark that our proposal in the $\mn$ can produce signals including up to six leptons plus MET. 
{In the region of the parameter space explored in the present work, the leptons not originated from the decay of a gauge boson will have too small transverse momentum due to the compressed nature of the spectrum. In this situation, it is unlikely that sufficient leptons would be detected to be able to use a high multiplicity of them as a powerful discriminant. On the contrary, the regions where the mass separation between neutralino-slepton/sneutrino is large enough are potentially testable requiring five or more leptons. Existing searches for events with four or more 
leptons~\cite{Aaboud:2018zeb} use a moderate associated MET or a large effective mass, to target the pair production of sleptons with masses above 500 GeV. These searches are insensitive to the small-mass slepton/sneutrino case that we were able to cover in this work. Nonetheless, the most important source of irreducible background to a signal with five or more leptons at $pp$ colliders is the triple-gauge boson production.
{Specifically the $ZZZ$ production cross section is 0.316 fb and the $ZZW$ one is 1.073 fb at $\sqrt{s}=13$ TeV, that means 0.07 expected background events including five or more leptons with an integrated luminosity of 36.1 fb$^{-1}$. In comparison, the production of slepton/sneutrino pairs with masses of 300 GeV will produce 18 events with five or more leptons with the same integrated luminosity.}
Thus, a dedicated experimental analysis using five or more leptons as the main discriminant might be sensitive to the slepton/sneutrino pair production in a larger mass region of the $\mn$. Although a similar final state is predicted assuming a single RPV channel $\lambda_{12k} \hat L_1 \hat L_2 \hat e^c_k$ in the superpotential, note however that the scenario of light sleptons/sneutrinos is already ruled out by current searches~\cite{Aaboud:2018zeb} unlike the case of the $\mn$.
}


\begin{acknowledgments}

The work of IL and CM was supported in part by the Spanish Agencia Estatal de Investigaci\'on 
through the grants 
FPA2015-65929-P (MINECO/FEDER, UE) and IFT Centro de Excelencia Severo Ochoa SEV-2016-0597. The work 
of DL was supported by the Argentinian CONICET, and also acknowledges the support of the Spanish grant FPA2015-65929-P (MINECO/FEDER, UE). 
The authors acknowledge the support of the Spanish Red Consolider MultiDark FPA2017-90566-REDC.
\end{acknowledgments}

\bibliographystyle{utphys}
\bibliography{munussm_v12}

\providecommand{\href}[2]{#2}\begingroup\raggedright\begin{thebibliography}{10}

\bibitem{Nilles:1983ge}
H.~P. Nilles, ``{Supersymmetry, Supergravity and Particle Physics},''
\href{http://dx.doi.org/10.1016/0370-1573(84)90008-5}{{\em Phys. Rept.}
  {\bfseries 110} (1984) 1--162}.

\bibitem{Haber:1984rc}
H.~E. Haber and G.~L. Kane, ``{The Search for Supersymmetry: Probing Physics
  Beyond the Standard Model},''
\href{http://dx.doi.org/10.1016/0370-1573(85)90051-1}{{\em Phys. Rept.}
  {\bfseries 117} (1985) 75--263}.

\bibitem{Martin:1997ns}
S.~P. Martin, ``{A Supersymmetry primer},''
  \href{http://arxiv.org/abs/hep-ph/9709356}{{\ttfamily arXiv:hep-ph/9709356
  [hep-ph]}}.
[Adv. Ser. Direct. High Energy Phys.18,1(1998)].

\bibitem{Tanabashi:2018oca}
{\bfseries Particle Data Group} Collaboration, M.~Tanabashi {\em et~al.},
  ``{Review of Particle Physics},''
\href{http://dx.doi.org/10.1103/PhysRevD.98.030001}{{\em Phys. Rev.} {\bfseries
  D98} no.~3, (2018) 030001}.

\bibitem{Barbier:2004ez}
R.~Barbier {\em et~al.}, ``{R-parity violating supersymmetry},''
  \href{http://dx.doi.org/10.1016/j.physrep.2005.08.006}{{\em Phys. Rept.}
  {\bfseries 420} (2005) 1--202},
\href{http://arxiv.org/abs/hep-ph/0406039}{{\ttfamily arXiv:hep-ph/0406039
  [hep-ph]}}.

\bibitem{LopezFogliani:2005yw}
D.~E. L\'opez-Fogliani and C.~Mu\~noz, ``{Proposal for a Supersymmetric
  Standard Model},''
  \href{http://dx.doi.org/10.1103/PhysRevLett.97.041801}{{\em Phys. Rev. Lett.}
  {\bfseries 97} (2006) 041801},
\href{http://arxiv.org/abs/hep-ph/0508297}{{\ttfamily arXiv:hep-ph/0508297
  [hep-ph]}}.

\bibitem{Escudero:2008jg}
N.~Escudero, D.~E. L\'opez-Fogliani, C.~Mu\~noz, and R.~Ruiz~de Austri,
  ``{Analysis of the parameter space and spectrum of the $\mu\nu$SSM},''
  \href{http://dx.doi.org/10.1088/1126-6708/2008/12/099}{{\em JHEP} {\bfseries
  12} (2008) 099},
\href{http://arxiv.org/abs/0810.1507}{{\ttfamily arXiv:0810.1507 [hep-ph]}}.

\bibitem{Ghosh:2017yeh}
P.~Ghosh, I.~Lara, D.~E. L\'opez-Fogliani, C.~Mu\~noz, and R.~Ruiz~de Austri,
  ``{Searching for left sneutrino LSP at the LHC},''
  \href{http://dx.doi.org/10.1142/S0217751X18501105}{{\em Int. J. Mod. Phys.}
  {\bfseries A33} no.~18n19, (2018) 1850110},
\href{http://arxiv.org/abs/1707.02471}{{\ttfamily arXiv:1707.02471 [hep-ph]}}.

\bibitem{Lara:2018rwv}
I.~Lara, D.~E. L\'opez-Fogliani, C.~Mu\~noz, N.~Nagata, H.~Otono, and R.~Ruiz
  De~Austri, ``{Looking for the left sneutrino LSP with displaced-vertex
  searches},'' \href{http://dx.doi.org/10.1103/PhysRevD.98.075004}{{\em Phys.
  Rev.} {\bfseries D98} no.~7, (2018) 075004},
\href{http://arxiv.org/abs/1804.00067}{{\ttfamily arXiv:1804.00067 [hep-ph]}}.

\bibitem{Ghosh:2008yh}
P.~Ghosh and S.~Roy, ``{Neutrino masses and mixing, lightest neutralino decays
  and a solution to the $\mu$ problem in supersymmetry},''
  \href{http://dx.doi.org/10.1088/1126-6708/2009/04/069}{{\em JHEP} {\bfseries
  04} (2009) 069},
\href{http://arxiv.org/abs/0812.0084}{{\ttfamily arXiv:0812.0084 [hep-ph]}}.

\bibitem{Bartl:2009an}
A.~Bartl, M.~Hirsch, A.~Vicente, S.~Liebler, and W.~Porod, ``{LHC phenomenology
  of the $\mu\nu$SSM},''
  \href{http://dx.doi.org/10.1088/1126-6708/2009/05/120}{{\em JHEP} {\bfseries
  05} (2009) 120},
\href{http://arxiv.org/abs/0903.3596}{{\ttfamily arXiv:0903.3596 [hep-ph]}}.

\bibitem{Ghosh:2012pq}
P.~Ghosh, D.~E. L\'opez-Fogliani, V.~A. Mitsou, C.~Mu\~noz, and R.~Ruiz~de
  Austri, ``{Probing the $\mu$-from-$\nu$ supersymmetric standard model with
  displaced multileptons from the decay of a Higgs boson at the LHC},''
  \href{http://dx.doi.org/10.1103/PhysRevD.88.015009}{{\em Phys. Rev.}
  {\bfseries D88} (2013) 015009},
\href{http://arxiv.org/abs/1211.3177}{{\ttfamily arXiv:1211.3177 [hep-ph]}}.

\bibitem{Ghosh:2014ida}
P.~Ghosh, D.~E. L\'opez-Fogliani, V.~A. Mitsou, C.~Mu\~noz, and R.~Ruiz~de
  Austri, ``{Probing the $\mu\nu$SSM with light scalars, pseudoscalars and
  neutralinos from the decay of a SM-like Higgs boson at the LHC},''
  \href{http://dx.doi.org/10.1007/JHEP11(2014)102}{{\em JHEP} {\bfseries 11}
  (2014) 102},
\href{http://arxiv.org/abs/1410.2070}{{\ttfamily arXiv:1410.2070 [hep-ph]}}.

\bibitem{Abreu:1999qz}
{\bfseries DELPHI} Collaboration, P.~Abreu {\em et~al.}, ``{Search for
  supersymmetry with R-parity violating L L anti-E couplings at S**(1/2) =
  183-GeV},''
\href{http://dx.doi.org/10.1007/s100520050720}{{\em Eur. Phys. J.} {\bfseries
  C13} (2000) 591}.

\bibitem{Abreu:2000pi}
{\bfseries DELPHI} Collaboration, P.~Abreu {\em et~al.}, ``{Search for SUSY
  with R-parity violating LL anti-E couplings at s**(1/2) = 189-GeV},''
  \href{http://dx.doi.org/10.1016/S0370-2693(00)00776-0}{{\em Phys. Lett.}
  {\bfseries B487} (2000) 36},
\href{http://arxiv.org/abs/hep-ex/0103006}{{\ttfamily arXiv:hep-ex/0103006
  [hep-ex]}}.

\bibitem{Achard:2001ek}
{\bfseries L3} Collaboration, P.~Achard {\em et~al.}, ``{Search for R parity
  violating decays of supersymmetric particles in $e^{+} e^{-}$ collisions at
  LEP},'' \href{http://dx.doi.org/10.1016/S0370-2693(01)01367-3}{{\em Phys.
  Lett.} {\bfseries B524} (2002) 65},
\href{http://arxiv.org/abs/hep-ex/0110057}{{\ttfamily arXiv:hep-ex/0110057
  [hep-ex]}}.

\bibitem{Heister:2002jc}
{\bfseries ALEPH} Collaboration, A.~Heister {\em et~al.}, ``{Search for
  supersymmetric particles with R parity violating decays in $e^{+} e^{-}$
  collisions at $\sqrt{s}$ up to 209-GeV},''
  \href{http://dx.doi.org/10.1140/epjc/s2003-01311-5}{{\em Eur. Phys. J.}
  {\bfseries C31} (2003) 1},
\href{http://arxiv.org/abs/hep-ex/0210014}{{\ttfamily arXiv:hep-ex/0210014
  [hep-ex]}}.

\bibitem{Abbiendi:2003rn}
{\bfseries OPAL} Collaboration, G.~Abbiendi {\em et~al.}, ``{Search for R
  parity violating decays of scalar fermions at LEP},''
  \href{http://dx.doi.org/10.1140/epjc/s2004-01596-8}{{\em Eur. Phys. J.}
  {\bfseries C33} (2004) 149},
\href{http://arxiv.org/abs/hep-ex/0310054}{{\ttfamily arXiv:hep-ex/0310054
  [hep-ex]}}.

\bibitem{Abdallah:2003xc}
{\bfseries DELPHI} Collaboration, J.~Abdallah {\em et~al.}, ``{Search for
  supersymmetric particles assuming R-parity nonconservation in e+ e-
  collisions at s**(1/2) = 192-GeV to 208-GeV},''
  \href{http://dx.doi.org/10.1140/epjc/s2004-01881-6,
  10.1140/epjc/s2004-01976-0}{{\em Eur. Phys. J.} {\bfseries C36} no.~1, (2004)
  1}, \href{http://arxiv.org/abs/hep-ex/0406009}{{\ttfamily
  arXiv:hep-ex/0406009 [hep-ex]}}.
[Erratum: Eur. Phys. J.C37,no.1,129(2004)].

\bibitem{Aad:2015rba}
{\bfseries ATLAS} Collaboration, G.~Aad {\em et~al.}, ``{Search for massive,
  long-lived particles using multitrack displaced vertices or displaced lepton
  pairs in pp collisions at $\sqrt{s}$ = 8 TeV with the ATLAS detector},''
  \href{http://dx.doi.org/10.1103/PhysRevD.92.072004}{{\em Phys. Rev.}
  {\bfseries D92} no.~7, (2015) 072004},
\href{http://arxiv.org/abs/1504.05162}{{\ttfamily arXiv:1504.05162 [hep-ex]}}.

\bibitem{Sirunyan:2017qaj}
{\bfseries CMS} Collaboration, A.~M. Sirunyan {\em et~al.}, ``{Search for new
  phenomena in final states with two opposite-charge, same-flavor leptons,
  jets, and missing transverse momentum in pp collisions at $ \sqrt{s}=13 $
  TeV},'' \href{http://dx.doi.org/10.1007/s13130-018-7845-2,
  10.1007/JHEP03(2018)076}{{\em JHEP} {\bfseries 03} (2018) 076},
\href{http://arxiv.org/abs/1709.08908}{{\ttfamily arXiv:1709.08908 [hep-ex]}}.

\bibitem{Aaboud:2018jiw}
{\bfseries ATLAS} Collaboration, M.~Aaboud {\em et~al.}, ``{Search for
  electroweak production of supersymmetric particles in final states with two
  or three leptons at $\sqrt{s}=13\,$TeV with the ATLAS detector},''
\href{http://arxiv.org/abs/1803.02762}{{\ttfamily arXiv:1803.02762 [hep-ex]}}.

\bibitem{Aaboud:2018zeb}
{\bfseries ATLAS} Collaboration, M.~Aaboud {\em et~al.}, ``{Search for
  supersymmetry in events with four or more leptons in $\sqrt{s}=13$ TeV $pp$
  collisions with ATLAS},''
  \href{http://dx.doi.org/10.1103/PhysRevD.98.032009}{{\em Phys. Rev.}
  {\bfseries D98} no.~3, (2018) 032009},
\href{http://arxiv.org/abs/1804.03602}{{\ttfamily arXiv:1804.03602 [hep-ex]}}.

\bibitem{Aaboud:2018sua}
{\bfseries ATLAS} Collaboration, M.~Aaboud {\em et~al.}, ``{Search for
  chargino-neutralino production using recursive jigsaw reconstruction in final
  states with two or three charged leptons in proton-proton collisions at
  $\sqrt{s}=13$ TeV with the ATLAS detector},''
\href{http://arxiv.org/abs/1806.02293}{{\ttfamily arXiv:1806.02293 [hep-ex]}}.

\bibitem{Athron:2018vxy}
{\bfseries GAMBIT} Collaboration, P.~Athron {\em et~al.}, ``{Combined collider
  constraints on neutralinos and charginos},''
\href{http://arxiv.org/abs/1809.02097}{{\ttfamily arXiv:1809.02097 [hep-ph]}}.

\bibitem{Carena:2018nlf}
M.~Carena, J.~Osborne, N.~R. Shah, and C.~E.~M. Wagner, ``{Supersymmetry and
  LHC Missing Energy Signals},''
\href{http://arxiv.org/abs/1809.11082}{{\ttfamily arXiv:1809.11082 [hep-ph]}}.

\bibitem{Fidalgo:2009dm}
J.~Fidalgo, D.~E. Lopez-Fogliani, C.~Munoz, and R.~Ruiz~de Austri, ``{Neutrino
  Physics and Spontaneous CP Violation in the $\mu\nu$SSM},''
  \href{http://dx.doi.org/10.1088/1126-6708/2009/08/105}{{\em JHEP} {\bfseries
  08} (2009) 105},
\href{http://arxiv.org/abs/0904.3112}{{\ttfamily arXiv:0904.3112 [hep-ph]}}.

\bibitem{Ghosh:2010zi}
P.~Ghosh, P.~Dey, B.~Mukhopadhyaya, and S.~Roy, ``{Radiative contribution to
  neutrino masses and mixing in $\mu\nu$SSM},''
  \href{http://dx.doi.org/10.1007/JHEP05(2010)087}{{\em JHEP} {\bfseries 05}
  (2010) 087},
\href{http://arxiv.org/abs/1002.2705}{{\ttfamily arXiv:1002.2705 [hep-ph]}}.

\bibitem{Ellwanger:2009dp}
U.~Ellwanger, C.~Hugonie, and A.~M. Teixeira, ``{The Next-to-Minimal
  Supersymmetric Standard Model},''
  \href{http://dx.doi.org/10.1016/j.physrep.2010.07.001}{{\em Phys. Rept.}
  {\bfseries 496} (2010) 1--77},
\href{http://arxiv.org/abs/0910.1785}{{\ttfamily arXiv:0910.1785 [hep-ph]}}.

\bibitem{Dawson:1983fw}
S.~Dawson, E.~Eichten, and C.~Quigg, ``{Search for Supersymmetric Particles in
  Hadron - Hadron Collisions},''
\href{http://dx.doi.org/10.1103/PhysRevD.31.1581}{{\em Phys. Rev.} {\bfseries
  D31} (1985) 1581}.

\bibitem{Eichten:1984eu}
E.~Eichten, I.~Hinchliffe, K.~D. Lane, and C.~Quigg, ``{Super Collider
  Physics},'' \href{http://dx.doi.org/10.1103/RevModPhys.56.579,
  10.1103/RevModPhys.58.1065}{{\em Rev. Mod. Phys.} {\bfseries 56} (1984) 579}.
[Addendum: Rev. Mod. Phys.58,1065(1986)].

\bibitem{delAguila:1990yw}
F.~del Aguila and L.~Ametller, ``{On the detectability of sleptons at large
  hadron colliders},''
\href{http://dx.doi.org/10.1016/0370-2693(91)90336-O}{{\em Phys. Lett.}
  {\bfseries B261} (1991) 326}.

\bibitem{Baer:1993ew}
H.~Baer, C.-h. Chen, F.~Paige, and X.~Tata, ``{Detecting Sleptons at Hadron
  Colliders and Supercolliders},''
  \href{http://dx.doi.org/10.1103/PhysRevD.49.3283}{{\em Phys. Rev.} {\bfseries
  D49} (1994) 3283},
\href{http://arxiv.org/abs/hep-ph/9311248}{{\ttfamily arXiv:hep-ph/9311248
  [hep-ph]}}.

\bibitem{Baer:1997nh}
H.~Baer, B.~W. Harris, and M.~H. Reno, ``{Next-to-leading order slepton pair
  production at hadron colliders},''
  \href{http://dx.doi.org/10.1103/PhysRevD.57.5871}{{\em Phys. Rev.} {\bfseries
  D57} (1998) 5871},
\href{http://arxiv.org/abs/hep-ph/9712315}{{\ttfamily arXiv:hep-ph/9712315
  [hep-ph]}}.

\bibitem{Bozzi:2004qq}
G.~Bozzi, B.~Fuks, and M.~Klasen, ``{Slepton production in polarized hadron
  collisions},'' \href{http://dx.doi.org/10.1016/j.physletb.2005.01.060}{{\em
  Phys. Lett.} {\bfseries B609} (2005) 339},
\href{http://arxiv.org/abs/hep-ph/0411318}{{\ttfamily arXiv:hep-ph/0411318
  [hep-ph]}}.

\bibitem{Aad:2014iza}
{\bfseries ATLAS} Collaboration, G.~Aad {\em et~al.}, ``{Search for
  supersymmetry in events with four or more leptons in $\sqrt{s}$ = 8 TeV pp
  collisions with the ATLAS detector},''
  \href{http://dx.doi.org/10.1103/PhysRevD.90.052001}{{\em Phys. Rev.}
  {\bfseries D90} no.~5, (2014) 052001},
\href{http://arxiv.org/abs/1405.5086}{{\ttfamily arXiv:1405.5086 [hep-ex]}}.

\bibitem{Fuks:2012qx}
B.~Fuks, M.~Klasen, D.~R. Lamprea, and M.~Rothering, ``{Gaugino production in
  proton-proton collisions at a center-of-mass energy of 8 TeV},''
  \href{http://dx.doi.org/10.1007/JHEP10(2012)081}{{\em JHEP} {\bfseries 10}
  (2012) 081},
\href{http://arxiv.org/abs/1207.2159}{{\ttfamily arXiv:1207.2159 [hep-ph]}}.

\bibitem{Fuks:2013vua}
B.~Fuks, M.~Klasen, D.~R. Lamprea, and M.~Rothering, ``{Precision predictions
  for electroweak superpartner production at hadron colliders with {\sc
  Resummino}},'' \href{http://dx.doi.org/10.1140/epjc/s10052-013-2480-0}{{\em
  Eur. Phys. J. C} {\bfseries 73} (2013) 2480},
\href{http://arxiv.org/abs/1304.0790}{{\ttfamily arXiv:1304.0790 [hep-ph]}}.

\bibitem{Fuks:2013lya}
B.~Fuks, M.~Klasen, D.~R. Lamprea, and M.~Rothering, ``{Revisiting slepton pair
  production at the Large Hadron Collider},''
  \href{http://dx.doi.org/10.1007/JHEP01(2014)168}{{\em JHEP} {\bfseries 01}
  (2014) 168},
\href{http://arxiv.org/abs/1310.2621}{{\ttfamily arXiv:1310.2621}}.

\bibitem{Jackson:2016mfb}
P.~Jackson, C.~Rogan, and M.~Santoni, ``{Sparticles in motion: Analyzing
  compressed SUSY scenarios with a new method of event reconstruction},''
  \href{http://dx.doi.org/10.1103/PhysRevD.95.035031}{{\em Phys. Rev.}
  {\bfseries D95} no.~3, (2017) 035031},
\href{http://arxiv.org/abs/1607.08307}{{\ttfamily arXiv:1607.08307 [hep-ph]}}.

\bibitem{Jackson:2017gcy}
P.~Jackson and C.~Rogan, ``{Recursive Jigsaw Reconstruction: HEP event analysis
  in the presence of kinematic and combinatoric ambiguities},''
  \href{http://dx.doi.org/10.1103/PhysRevD.96.112007}{{\em Phys. Rev.}
  {\bfseries D96} no.~11, (2017) 112007},
\href{http://arxiv.org/abs/1705.10733}{{\ttfamily arXiv:1705.10733 [hep-ph]}}.

\bibitem{Conte:2012fm}
E.~Conte, B.~Fuks, and G.~Serret, ``{MadAnalysis 5, A User-Friendly Framework
  for Collider Phenomenology},''
  \href{http://dx.doi.org/10.1016/j.cpc.2012.09.009}{{\em Comput. Phys.
  Commun.} {\bfseries 184} (2013) 222--256},
\href{http://arxiv.org/abs/1206.1599}{{\ttfamily arXiv:1206.1599 [hep-ph]}}.

\bibitem{Conte:2014zja}
E.~Conte, B.~Dumont, B.~Fuks, and C.~Wymant, ``{Designing and recasting LHC
  analyses with MadAnalysis 5},''
  \href{http://dx.doi.org/10.1140/epjc/s10052-014-3103-0}{{\em Eur. Phys. J.}
  {\bfseries C74} no.~10, (2014) 3103},
\href{http://arxiv.org/abs/1405.3982}{{\ttfamily arXiv:1405.3982 [hep-ph]}}.

\bibitem{Dumont:2014tja}
B.~Dumont, B.~Fuks, S.~Kraml, S.~Bein, G.~Chalons, E.~Conte, S.~Kulkarni,
  D.~Sengupta, and C.~Wymant, ``{Toward a public analysis database for LHC new
  physics searches using MADANALYSIS 5},''
  \href{http://dx.doi.org/10.1140/epjc/s10052-014-3242-3}{{\em Eur. Phys. J.}
  {\bfseries C75} no.~2, (2015) 56},
\href{http://arxiv.org/abs/1407.3278}{{\ttfamily arXiv:1407.3278 [hep-ph]}}.

\bibitem{Alwall:2014hca}
J.~Alwall, R.~Frederix, S.~Frixione, V.~Hirschi, F.~Maltoni, O.~Mattelaer,
  H.~S. Shao, T.~Stelzer, P.~Torrielli, and M.~Zaro, ``{The automated
  computation of tree-level and next-to-leading order differential cross
  sections, and their matching to parton shower simulations},''
  \href{http://dx.doi.org/10.1007/JHEP07(2014)079}{{\em JHEP} {\bfseries 07}
  (2014) 079},
\href{http://arxiv.org/abs/1405.0301}{{\ttfamily arXiv:1405.0301 [hep-ph]}}.

\bibitem{Sjostrand:2006za}
T.~Sjostrand, S.~Mrenna, and P.~Z. Skands, ``{PYTHIA 6.4 physics and manual},''
  \href{http://dx.doi.org/10.1088/1126-6708/2006/05/026}{{\em JHEP} {\bfseries
  05} (2006) 026},
\href{http://arxiv.org/abs/hep-ph/0603175}{{\ttfamily arXiv:hep-ph/0603175
  [hep-ph]}}.

\bibitem{ATL-PHYS-PUB-2014-021}
``{ATLAS Run 1 Pythia8 tunes},'' Tech. Rep. ATL-PHYS-PUB-2014-021, CERN,
  Geneva, Nov, 2014.
\newblock \url{https://cds.cern.ch/record/1966419}.

\bibitem{deFavereau:2013fsa}
{\bfseries DELPHES 3} Collaboration, J.~de~Favereau, C.~Delaere, P.~Demin,
  A.~Giammanco, V.~Lemaitre, A.~Mertens, and M.~Selvaggi, ``{DELPHES 3, A
  modular framework for fast simulation of a generic collider experiment},''
  \href{http://dx.doi.org/10.1007/JHEP02(2014)057}{{\em JHEP} {\bfseries 02}
  (2014) 057},
\href{http://arxiv.org/abs/1307.6346}{{\ttfamily arXiv:1307.6346 [hep-ex]}}.

\bibitem{Staub:2008uz}
F.~Staub, ``{SARAH},''
\href{http://arxiv.org/abs/0806.0538}{{\ttfamily arXiv:0806.0538 [hep-ph]}}.

\bibitem{Staub:2011dp}
F.~Staub, T.~Ohl, W.~Porod, and C.~Speckner, ``{A Tool Box for Implementing
  Supersymmetric Models},''
  \href{http://dx.doi.org/10.1016/j.cpc.2012.04.013}{{\em Comput. Phys.
  Commun.} {\bfseries 183} (2012) 2165--2206},
\href{http://arxiv.org/abs/1109.5147}{{\ttfamily arXiv:1109.5147 [hep-ph]}}.

\bibitem{Staub:2013tta}
F.~Staub, ``{SARAH 4 : A tool for (not only SUSY) model builders},''
  \href{http://dx.doi.org/10.1016/j.cpc.2014.02.018}{{\em Comput. Phys.
  Commun.} {\bfseries 185} (2014) 1773},
\href{http://arxiv.org/abs/1309.7223}{{\ttfamily arXiv:1309.7223 [hep-ph]}}.

\bibitem{Porod:2003um}
W.~Porod, ``{SPheno, a program for calculating supersymmetric spectra, SUSY
  particle decays and SUSY particle production at e+ e- colliders},''
  \href{http://dx.doi.org/10.1016/S0010-4655(03)00222-4}{{\em Comput. Phys.
  Commun.} {\bfseries 153} (2003) 275},
\href{http://arxiv.org/abs/hep-ph/0301101}{{\ttfamily arXiv:hep-ph/0301101
  [hep-ph]}}.

\bibitem{Porod:2011nf}
W.~Porod and F.~Staub, ``{SPheno 3.1: Extensions including flavour, CP-phases
  and models beyond the MSSM},''
  \href{http://dx.doi.org/10.1016/j.cpc.2012.05.021}{{\em Comput. Phys.
  Commun.} {\bfseries 183} (2012) 2458},
\href{http://arxiv.org/abs/1104.1573}{{\ttfamily arXiv:1104.1573 [hep-ph]}}.

\bibitem{Buckley:2014ana}
A.~Buckley, J.~Ferrando, S.~Lloyd, K.~Nordstrom, B.~Page, M.~Rufenacht,
  M.~Schonherr, and G.~Watt, ``{LHAPDF6: parton density access in the LHC
  precision era},''
  \href{http://dx.doi.org/10.1140/epjc/s10052-015-3318-8}{{\em Eur. Phys. J.}
  {\bfseries C75} (2015) 132},
\href{http://arxiv.org/abs/1412.7420}{{\ttfamily arXiv:1412.7420 [hep-ph]}}.

\bibitem{Bozzi:2006fw}
G.~Bozzi, B.~Fuks, and M.~Klasen, ``{Transverse-momentum resummation for
  slepton-pair production at the CERN LHC},''
  \href{http://dx.doi.org/10.1103/PhysRevD.74.015001}{{\em Phys. Rev.}
  {\bfseries D74} (2006) 015001},
\href{http://arxiv.org/abs/hep-ph/0603074}{{\ttfamily arXiv:hep-ph/0603074
  [hep-ph]}}.

\bibitem{Bozzi:2007qr}
G.~Bozzi, B.~Fuks, and M.~Klasen, ``{Threshold Resummation for Slepton-Pair
  Production at Hadron Colliders},''
  \href{http://dx.doi.org/10.1016/j.nuclphysb.2007.03.052}{{\em Nucl. Phys.}
  {\bfseries B777} (2007) 157--181},
\href{http://arxiv.org/abs/hep-ph/0701202}{{\ttfamily arXiv:hep-ph/0701202
  [hep-ph]}}.

\bibitem{Bozzi:2007tea}
G.~Bozzi, B.~Fuks, and M.~Klasen, ``{Joint resummation for slepton pair
  production at hadron colliders},''
  \href{http://dx.doi.org/10.1016/j.nuclphysb.2007.10.021}{{\em Nucl. Phys.}
  {\bfseries B794} (2008) 46--60},
\href{http://arxiv.org/abs/0709.3057}{{\ttfamily arXiv:0709.3057 [hep-ph]}}.

\bibitem{An:2015rpe}
{\bfseries Daya Bay} Collaboration, F.~P. An {\em et~al.}, ``{New measurement
  of antineutrino oscillation with the full detector configuration at Daya
  Bay},'' \href{http://dx.doi.org/10.1103/PhysRevLett.115.111802}{{\em Phys.
  Rev. Lett.} {\bfseries 115} no.~11, (2015) 111802},
\href{http://arxiv.org/abs/1505.03456}{{\ttfamily arXiv:1505.03456 [hep-ex]}}.

\bibitem{Ade:2015xua}
{\bfseries Planck} Collaboration, P.~A.~R. Ade {\em et~al.}, ``{Planck 2015
  results. XIII. Cosmological parameters},''
  \href{http://dx.doi.org/10.1051/0004-6361/201525830}{{\em Astron. Astrophys.}
  {\bfseries 594} (2016) A13},
\href{http://arxiv.org/abs/1502.01589}{{\ttfamily arXiv:1502.01589
  [astro-ph.CO]}}.

\bibitem{prepar}
I.~Lara, D.~E. L\'opez-Fogliani, and C.~Mu\~noz \href{http://arxiv.org/abs/in
  preparation}{{\ttfamily in preparation}}.

\end{thebibliography}\endgroup
\end{document}